\documentclass[fleqn,a4paper]{vch-book}
\usepackage{amsmath}
\usepackage{amssymb}
\usepackage{makeidx}
    \makeindex
\usepackage{times}
\usepackage{tabularx}

\usepackage[dvips]{epsfig}
\DeclareGraphicsExtensions{.eps,.ps}

\chardef\bslash=`\\ 

\newcommand{\bibtex}{\ifx\is@itshape\f@shape{\fontshape{scit}\selectfont
Bib}\else\textsc{Bib}\fi\kern-.1em\TeX}

\newcommand{\beq}{\begin{equation}}
\newcommand{\eeq}{\end{equation}}
\newcommand{\bea}{\begin{eqnarray}}
\newcommand{\eea}{\end{eqnarray}}
\newcommand{\ba}{\begin{array}}
\newcommand{\ea}{\end{array}}
\newcommand{\bc}{\begin{center}}
\newcommand{\ec}{\end{center}}

\newcommand{\gsimeq}{\stackrel{>}{\scriptstyle\sim}}

\newcommand{\bml}{\begin{mathletters}}
\newcommand{\eml}{\end{mathletters}}
\newcommand{\commentout}[1]{{}}

\newcommand{\p}{{\bf p}}

\newcommand{\rb}{$^{85}$Rb }
\newcommand{\etal}{{\it et al.}}
\newcommand{\eq}[1]{(\ref{#1})}

\newcommand{\phidag}{\phi^\dagger}
\newcommand{\psidag}{\psi^\dagger}
\newcommand{\prl}{Phys. Rev. Lett. }
\newcommand{\pra}{Phys. Rev. A }
\newcommand{\rmp}{Rev. Mod. Phys. }

\begin{document}





\chapter{Role of Molecular Dissociation in Feshbach-Interacting
\rb Condensates \tocauthor{Matt Mackie,$^1$ Kalle-Antti
Suominen,$^{1,2}$ and Juha Javanainen$^3$}}\label{chap:role_mol_diss}

\authorafterheading{Matt Mackie,$^1$ Kalle-Antti
Suominen,$^{1,2}$ and Juha Javanainen$^3$}

\affil{$^1$Helsinki Institute of Physics, PL 64,
FIN-00014
Helsingin yliopisto, Finland \\
$^2$Department of Physics, University of
Turku, FIN-20014 Turun yliopisto, Finland \\
$^3$Physics Department, University of Connecticut,
Storrs,
Connecticut, 06269-3046, USA}

\section*{Abstract}
Recent Feshbach-resonance experiments with \rb
Bose-Einstein
condensates have led to a host of unexplained results:
dramatic
losses of condensate atoms for an across-resonance
sweep of the
magnetic field, a collapsing condensate with a burst
of
atoms emanating from the remnant condensate, increased
losses
for decreasing interaction times---until very short
times are reached, and
coherent oscillations between remnant and burst
atoms. In particular, the amplitude of the remnant-burst oscillations,
and the corresponding missing atoms, have prompted speculation as
to the formation of a molecular condensate. Using a minimal mean-field
model, we find that rogue dissociation, molecular dissociation to
noncondensate atom pairs, is qualitatively implicated as the physical
mechanism responsible for these observations, although very little
molecular condensate is formed. Refining the model provides excellent
quantitative agreement with the experimental remnant-burst oscillations,
and the fraction of molecular condensate accounts almost
entirely for the measured atom loss.

\section*{Introduction}
The process known as the Feshbach
resonance \cite{FES92} occurs when two ultracold atoms collide in
the
presence of a magnetic field, whereby a spin flip of
one atom
can induce the pair to jump from the two-atom
continuum to a
quasibound molecular state. If the initial atoms are
Bose condensed \cite{BEC}, the so-formed molecules
will also
comprise a Bose-Einstein
condensate~(BEC) \cite{MBEC_FR}. Since the
Feshbach resonance is mathematically identical to
photoassociation \cite{DRU98,JAV99,KOS00}, the process
that occurs when
two ultracold atoms form a molecule by absorbing a
photon, insight
gathered in either case is applicable to the other. In
particular, the most recent results from
photoassociation predict that
rogue dissociation, molecular dissociation to
noncondensate atom pairs,
imposes a maximum achievable rate on atom-molecule
conversion, as well as
the possibility of coherent Rabi oscillations between
the BEC
and dissociated atom pairs \cite{JAV02}.

Initial Feshbach resonance experiments in
\rb \cite{INIT_FRE} were motivated by the
possibility \cite{SHIFT} of enabling Bose condensation
by
tuning the natively negative atom-atom scattering
length into
the positive regime. As part of the experiment that
achieved
condensation \cite{COR00}, the magnetic field was
swept across
the Feshbach resonance, resulting in heavy
condensate losses ($\sim80$\% for the slowest sweep
rates).
Additional experiments led to the observation of
collapsing
condensates, an event characterized by bursts of atoms
emanating from a remnant BEC, and coined
''bosenova" for the
analogy with a supernova explosion \cite{DON01}. More recently,
experiments with pulsed magnetic fields that come
close to, but
do not cross, the Feshbach-resonance have revealed a
striking increase in condensate loss for a decrease in
the
interaction time-- until reaching very short times \cite{CLA02}. Finally,
double-pulse results
indicate large amplitude ($\sim 25\%$) remnant-burst oscillations, with
the missing atoms ($\sim 10\%$) prompting speculation on the
formation of a
molecular condensate \cite{DON02}.

In the mean time, work on Feshbach-stimulated
photoproduction of stable molecular condensates
indicates that rogue dissociation dominates
atom-molecule
conversion for the above \rb Feshbach
resonance, meaning that the production of a significant
fraction of molecular BEC is not to be expected \cite{MAC02a}.
Additionally, we find intriguing the assertions of a breakdown of
mean-field theory in the face of large resonance-induced scattering
lengths \cite{COR00,CLA02}, especially given that theory
actually faults the effective all-atom
description \cite{MFT_VALID}. We have therefore tested a
mean-field model of coherent atom-molecule conversion
against the salient
features of the JILA experiments \cite{MAC02b}, and the subsequent
understanding is presented in this Contribution (for
related work, see Refs. \cite{RELATEDa,RELATEDb}).

\section*{Mean-field theory and its validity}
The mathematical equivalence
of the
Feshbach-resonance \cite{MBEC_FR} and
photoassociation \cite{DRU98,JAV99}
lies in both processes being described, in
terms of second
quantization, as destroying two atoms and creating a
molecule. We
therefore model a quantum degenerate gas of atoms
coupled via a Feshbach
resonance to a condensate of quasibound molecules
based on
Refs. \cite{JAV99,KOS00,JAV02}. The initial atoms are
denoted by the
boson field
$\phi({\bf r},t)$, and the quasibound molecules by the
field
$\psi({\bf r},t)$. The Hamiltonian density for this system
is
\bea \frac{\cal H}{\hbar} &=&
\phidag\left[-\frac{\hbar\nabla^2}{2m}\right]\phi\nonumber
+\psidag\left[-\frac{\hbar\nabla^2}{4m}+\delta_0\right]\psi
\\&& -\frac{\Omega}{2\sqrt{\rho}}\,[\psi^\dagger\phi\phi
+\phi^\dagger\phi^\dagger\psi]
+\frac{2\pi\hbar a}{m}\,\phi^\dagger\phi^\dagger\phi\phi\,,
\label{CURHAM}
\eea
\beq
\Omega =
\lim_{\epsilon\rightarrow0}\sqrt{\frac{\sqrt2\pi\hbar^{3/2}\rho}
{\mu^{3/2}}\frac{\Gamma(\epsilon)}{\sqrt{\epsilon}}}\,,
\label{OMDEF}
\eeq
where $m=2\mu$ is the mass of an atom, $\hbar\delta_0$
is the energy
difference between a
molecule and two atoms, $a$ is the off-resonant $s$-wave scattering
length,
$\rho$ is an invariant density equal to the sum of atom density and twice
the molecular
density, and
$\Gamma(\epsilon)$ is the dissociation rate for a
molecule with
the energy
$\hbar\epsilon$ above the threshold of the
Feshbach resonance.

To address the validity of mean-field theory for
near-resonant systems, we find the
Heisenberg equation of motion for the molecular operator and solve it
in the adiabatic limit:
$
\psi = [\Omega/ (2\delta_0\sqrt{\rho})]\,\phi\phi\,.
$
Pluging this back into Eq.~\eq{CURHAM}, gives
\beq
\frac{\cal H}{\hbar} = \frac{2\pi\hbar a_{eff}}{m}\,
\phi^\dagger\phi^\dagger
\phi\phi\,;\quad
a_{eff} = a\left[1-\frac{m\Omega^2}{4\pi\hbar\delta_0\rho a} \right]\,.
\label{VARSCL}
\eeq
The detuning can of course written in terms of the magnetic field by
introducing the difference in the magnetic  moments
of the resonant states $\Delta_\mu\,$:
$
\hbar\delta_0 = \Delta_\mu (B-B_0).
$
Similarly, the condensate coupling can be approximated as
$\Omega=[2\pi\rho |a|\mu_{ma}\Delta_R/m]^{1/2}$. Substituting these
expressions leads immediately to the standard way of writing the
scattering length around the Feshbach resonance:
\beq
a_{eff} = a\left( 1-\frac{\Delta_B}{B - B_0}\right)\,.
\label{SCLB}
\eeq

The effective scattering
length~(\ref{SCLB}),
was obtained by adiabatically eliminating the molecular field, i.e., by
assuming
$\delta_0\gg2\pi\hbar a/m,\Omega$. Since the adiabatic
approximation fails for $\delta_0\rightarrow0$ ($B\rightarrow B_0$),
it is therefore the effective all-atom description that actually breaks
down when resonance is encountered \cite{MFT_VALID}, {\em not} mean-field
theory. The validity of the mean field approximation is determined by
$\rho|a|^3\ll 1$, which holds independent of the value of the magnetic
field. Including the molecular condensate dynamics provides a complete
mean-field description of the near-resonant system.

That said, we expect
the mean-field equations arising from the Hamiltonian~(\ref{CURHAM}) to
suitably model the salient features of the JILA experiments. Switching to
momentum space, only zero-momentum atomic and molecular
condensate modes are
retained, represented by the respective $c$-number
amplitudes
$\alpha$ and
$\beta$. We also take into account correlated pairs of
noncondensate
atoms using a complex amplitude
$C(\epsilon)$, which represent pairs of noncondensate
atoms in the manner
of the Heisenberg picture expectation value $\langle
a_\p
a_{-\p}\rangle$, with $\hbar\epsilon$ being the
relative energy of the
atoms.  The normalization of our mean fields is such
that
$|\alpha|^2+|\beta|^2+\int d\epsilon\,|C(\epsilon)|^2
= 1$.  We work from
the Heisenberg equation of motion of the boson
operators under the
simplifying assumption that the noncondensate atoms
pairs are only
allowed to couple back to the molecular condensate,
ignoring the
possibility that noncondensate atoms associate to make
noncondensate molecules. This neglect
is justified to the extent that
Bose enhancement favors transitions back to the
molecular condensate. The
final mean-field equations are \cite{JAV02}
\bea
i\dot\alpha &=&
     -\frac{\Omega}{\sqrt{2}}\alpha^*\beta, \\
i\dot\beta &=& \delta_0\beta
-\frac{\Omega}{\sqrt{2}}\alpha\alpha
                  -\frac{\xi}{\sqrt{2\pi}}\int
d\epsilon\,\sqrt[4]{\epsilon}\,C(\epsilon),
\\
i\dot C(\epsilon) &=& \epsilon C(\epsilon)
-\frac{\xi}{\sqrt{2\pi}}\,\sqrt[4]{\epsilon}\,\beta\,.
\eea
\label{EQM}
The analog of the Rabi frequency for the rogue
modes $\xi$ is inferred using Fermi Golden rule, which
gives the dissociation rate for a positive-energy
molecule as
$
\Gamma(\epsilon)=\sqrt{\epsilon}\,\xi^2\,.
$

Next the problem is reformulated in terms of two key
parameters
with the dimension of frequency. The density-dependent
frequency
$
\omega_\rho = {\hbar\rho^{2/3}/ m},
$
has been identified before, along with the
operational significance that, once
$\Omega\gsimeq\omega_\rho$, rogue dissociation is
expected to be a dominant factor in the
dynamics \cite{JAV99,KOS00,JAV02}. Here it is
convenient to define another primary parameter with
the dimension of frequency. Considering on-shell
dissociation of molecules to atoms with
the relative energy
$\epsilon$, the Wigner threshold law delivers a
dissociation rate
$\Gamma(\epsilon)$ such that
$\Gamma(\epsilon)/\sqrt\epsilon$ converges to a finite
limit for
$\epsilon\rightarrow0$; hence, we define
$
4\Xi = \left(\lim_{\epsilon\rightarrow 0}\,
{\Gamma(\epsilon)/
\sqrt\epsilon}\,
\right)^2,
$
which indeed has the dimension of frequency.
A combination of the preceeding equations gives the parameters in the
mean-field equations
as
\beq
\Omega = 2^{3/2}
\sqrt{\pi}\,\Xi^{1/4}\omega_\rho^{3/4}, \quad
\xi=\sqrt{2}\,\Xi^{1/4}\,.
\eeq

Lastly, when the coupling
to the continuum of
noncondensate atom pairs is included, the continuum
shifts the
molecular state \cite{CONT_SHIFT}. We have, of course, taken the
dominant state pushing, and the related renormalization effects,
into account in our calculations \cite{MAC02b,MAC03}. Suffice
it to say that we choose the bare detuning
$\delta_0$ so that the renormalized detuning attains the
desired value;
hereafter, we use the symbol $\delta = \Delta_\mu (B-B_0)/\hbar$
for the renormalized
detuning, which is the parameter that is varied experimentally by
changing the
laser frequency in photoassociation, or the magnetic
field in
the Feshbach resonance. The position
of the Feshbach
resonance is \cite{DON02} $B_0=154.9\,{\rm G}$, and the
difference in magnetic
moments between bound molecules and free atom pairs,
$\Delta_\mu\approx2\,\mu_B$
    (where $\mu_B$ is the Bohr magneton), is borrowed
from
${}^{87}$Rb \cite{WYN00}. We have estimated \cite{MAC02b}
$\Xi=5.29\times10^9\,{\rm s}^{-1}$, and
thus $\xi=381\,{\rm s}^{-1/4}$. Compared
to the ensuing detunings $\delta$, the interactions
energies
between the atoms due to the background scattering
length
$a=23.8\,{\rm nm}$ are immaterial. We therefore ignore
atom-atom
interactions unrelated to the Feshbach resonance, as
well as the
(unknown) atom-molecule and molecule-molecule
interactions.

\section*{Results}
\begin{figure}[htb]
\includegraphics[width=\textwidth]{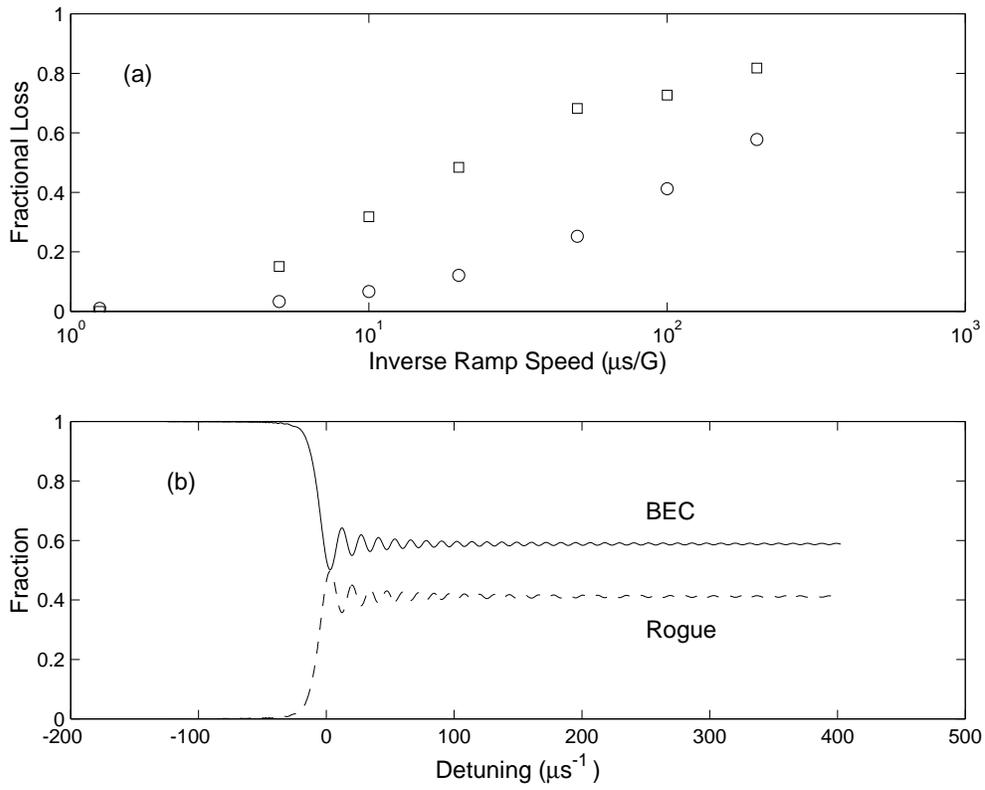}
	 \caption{(a) Experimental \protect\cite{COR00} and theoretical
($\circ$) atom loss incurred in sweeping a $^{85}$Rb
BEC across the
Feshbach resonance, where the magnetic field
is swept in a
linear fashion from $B_i=162$~G to $B_f=132$~G. In
each numerical run, the
fraction of molecular condensate is
$\sim 10^{-6}$. (b) Results for
$\dot{B}^{-1}=100\text{ $\mu$s/G}$ are typical, and
suggest that the
system undergoes collective adiabatic following from
BEC to dissociated
atom pairs.}
	 \label{SWEEP}
\end{figure}

We begin with the experiments \cite{COR00}
implementing a sweep of the magnetic field across the
Feshbach
resonance, which are of course a version of the
Landau-Zener
problem \cite{JAV99,KOS00,LZP}. Although a sweep of
the detuning
$\delta$ from above to below threshold at a rate slow
compared to
the condensate coupling $\Omega$ will move the system
adiabatically
from all atoms to all molecules, rogue dissociation
will overtake
coherent atom-molecule conversion when
$\Omega\gsimeq\omega_\rho$ \cite{JAV99,KOS00,JAV02}.
Nevermind that
the JILA experiments sweep from below to above
threshold, for a density
$\rho=1\times 10^{12}\,{\rm cm}^{-3}$ the condensate
coupling is
$\Omega=1.93\times10^5\,\text{s}^{-1}\approx
250\,\omega_\rho$, and so rogue dissociation should
seriously dominate. This is indeed the case
(see Fig.~\ref{SWEEP}). Apparently,
coherent conversion occurs
not between atomic and molecular BEC, but between
atomic BEC and
dissociated atom pairs. Holding this thought, we
conclude that mean-field
theory indicates rogue dissociation as a primary sink of
atoms in the Ref. \cite{COR00} sweeps across the
resonance.

\begin{figure}[htb]
\includegraphics[width=\textwidth]{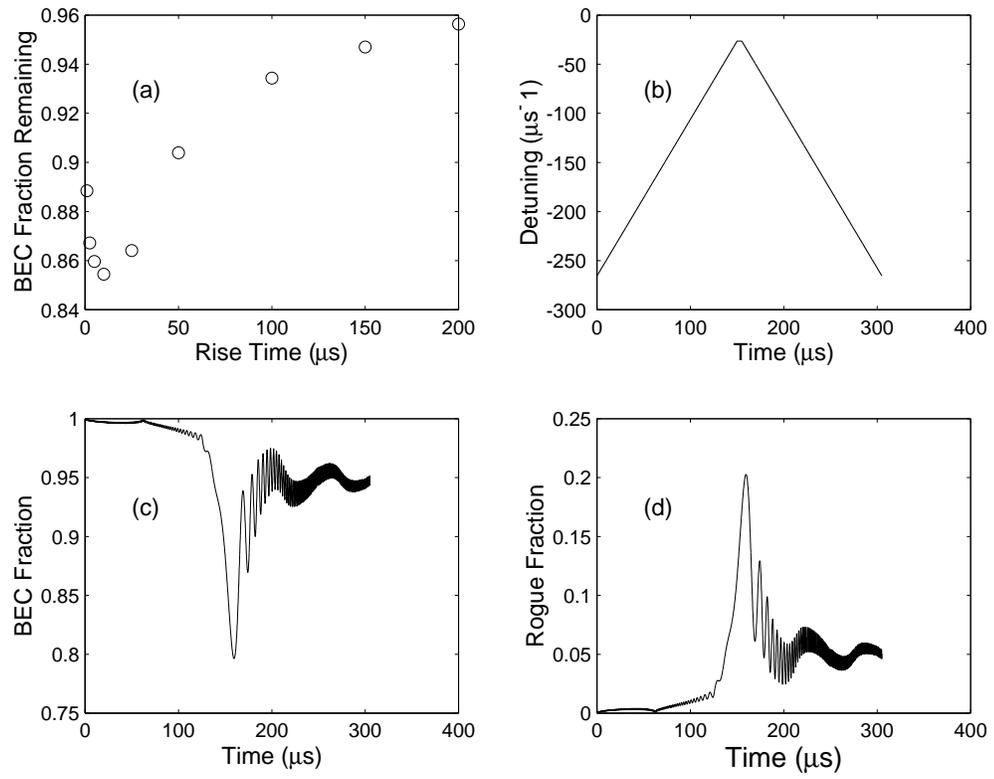}
	 \caption{Theory of a magnetic
field pulse applied
to a \rb condensate for
$\rho=1.9\times10^{13}\,\text{cm}^{-3}$ and
$\Omega=8.42\times10^5\,\text{s}^{-1}$. (a) Remnant
fraction
versus detuning (magnetic field) rise time.
(b-d) Results for a pulse with
$150\,\mu$s rise time indicate adiabatic passage of
BEC atoms to
and from dissociated atom pairs. The minimum in
panel (a), similar to Ref. \protect\cite{CLA02}, occurs at the onset of
adiabaticity.}
	 \label{PULSE}
\end{figure}

Next we consider the experiments \cite{CLA02} for which
nontrivial electromagnetic coil technology was
developed to create
trapezoidal magnetic field pulses that bring the
system near-- but
not across-- resonance, hold for a given amount of
time,
and return to the
original field value. Neglecting the burst, these
remnant-focused
experiments revealed a contradiction with the
conventional understanding
of condensate loss: rather than a loss that increased
monotonically with
increasing interaction time, the results
indicated a loss that
increased with {\em decreasing} interaction
time, until very short times were reached.
The present mean-field approach works similarly,
as shown in Fig.~\ref{PULSE}. Our interpretation is
that
adiabaticity is again at play. At very short
pulse durations, increasing interaction time leads to
increasing condensate loss, as expected. In contrast,
as the time dependence of the pulse gets slower, the
system eventually follows the pulse adiabatically,
and returns close to the initial condensate state
when the pulse has passed.

\begin{figure}[htb]
\includegraphics[width=\textwidth]{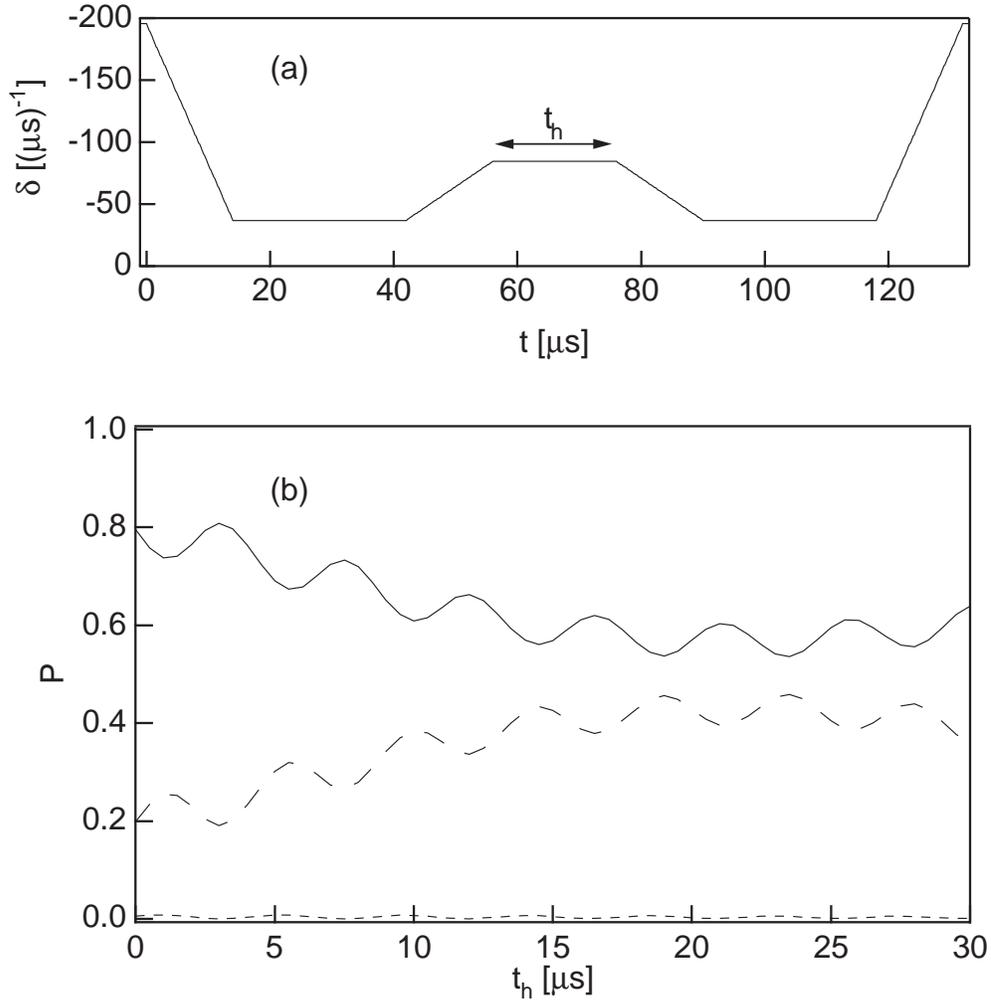}
	 \caption{Simulation of the Ref. \protect\cite{DON02}
experiments for a
density
$\rho=5.4\times10^{13}\text{cm}^{-3}$
and $\Omega=$ $1.42\times10^6\text{s}^{-1}$. (a) Time
dependence of the
detuning, and (b) the fraction of atoms in the remnant
condensate (solid
line), in noncondensate atoms pairs (dashed line) and
in the molecular
condensate (short-dashed line) after the pulse
sequence
as a function of the
hold time $t_h$ between the two trapezoidal pulses.
The frequency of the
oscillations is compatible with our predictions for the molecular
dissociation energy \protect\cite{MAC02b,MAC03}, identifying these
oscillations as Ramsey fringes in the transition between the
atomic condensate and a molecular condensate dressed by
dissociated atom pairs.}
	 \label{FRINGE}
\end{figure}

Finally, we turn to the experiments \cite{DON02} in which two
trapezoidal pulses were
applied to a \rb condensate, and the fraction of
remnant and burst
atoms measured for a variable between-pulse time and
magnetic-field amplitude. These
experiments revealed coherent remnant-burst
oscillations with amplitudes
of up to
$\sim 25$\%. As it happens, we have recently predicted
coherent
oscillations between atoms and dissociated atom pairs
in a
rogue-dominated system, although we harbored doubts
regarding any
practical realization \cite{JAV02}. Casting these
doubts aside, we
consider a time dependent detuning (magnetic field) similar to Fig.~2 of
Ref. \cite{DON02} [Fig.~\ref{FRINGE}(a)], and determine
the fraction of remnant condensate atoms,
noncondensate atoms, and
molecules at the end of the pulse sequence as a
function of the holding
time between the two pulses [Fig.~\ref{FRINGE}(b)].
Oscillations are seen with the amplitude of about 15\%
between condensate
and noncondensate atoms at the frequency of the
molecular state
corresponding to the magnetic field during the holding
period. The
molecular fraction appears too small to account for
the amplitude of the
oscillations. In fact, what we termed molecular
frequency is the
characteristic frequency of a coherent superposition
of
molecules and
noncondensate atom pair. Here the oscillations,
directly
comparable to Fig.~4(a) in Ref. \cite{DON02}, are
Ramsey
fringes \cite{RAM50} in the evolution between an
atomic condensate
and a molecular condensate dressed with noncondensate
atom pairs.

\begin{figure}[htb]
\includegraphics[width=\textwidth]{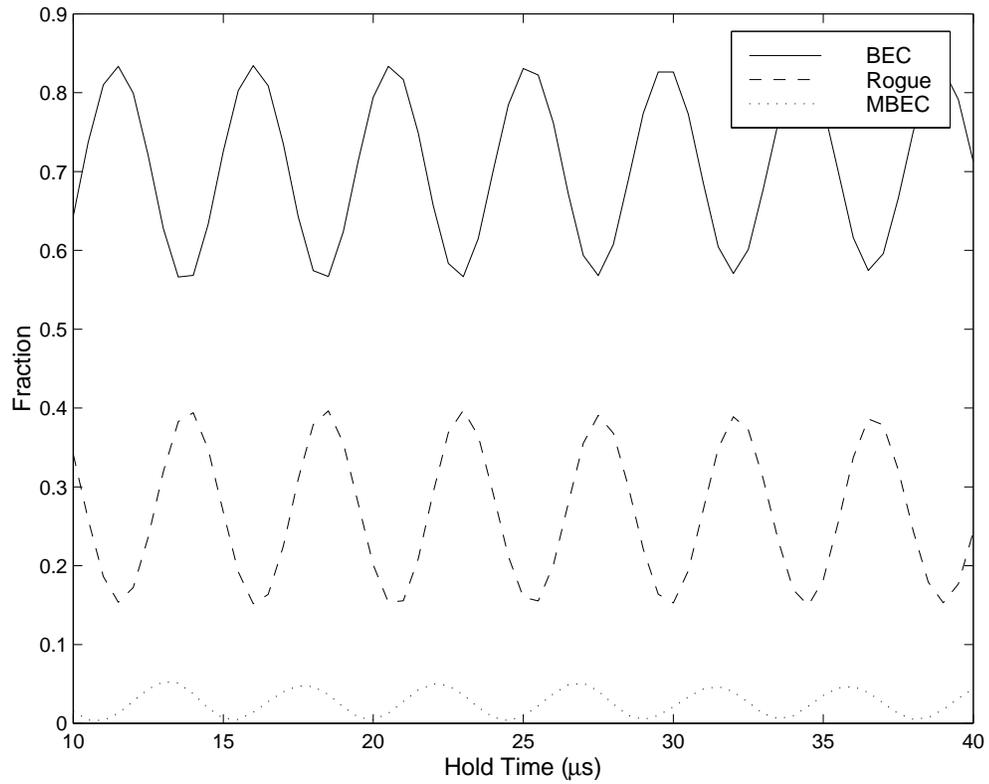}
	 \caption{Full-model \protect\cite{MAC03} simulation of the
Ref. \protect\cite{DON02} experiments for a peak
density
$\rho_0=1.1\times10^{13}\text{cm}^{-3}$. Fraction of atoms in the
remnant condensate (solid
line), in noncondensate atoms pairs (dashed line) and
in the molecular
condensate (dotted line) after the pulse
sequence
as a function of the
hold time between the two trapezoidal pulses. The magnetic field pulse is
similar to that shown in Fig. \protect\ref{FRINGE}. Not only does the
fringe amplitude agree quantitatively with the experimental observation,
but the fraction of molecules formed is entirely consistent with the
measured atom loss (see Fig. 6 of Ref. \protect\cite{DON02}).}
	 \label{NEW_FRINGE}
\end{figure}

Although our rogue-dissociation ideas provide a neat
qualitative explanation for the three experiments we
have discussed\footnote{If an explanation in terms of below-threshold
molecular dissociation seems a bit odd, consider that energy need not be
conserved for transient processes where a time dependence is externally
imposed on the system.}, in all fairness it must
be noted
that we have fallen short of a full quantitative
agreement. We have therefore refined our renormalization techniques,
extended our model to allow for Bose enhancement of the rogue
modes, and included an average over a Gaussian distribution
of densities \cite{MAC03}. So far, we have only applied this full model
to the double-pulse experiments \cite{DON02}, the results of which
are shown in Fig.~\ref{NEW_FRINGE}. Not only do we find excellent
agreement on the size of the experimental Ramsey fringes, but
the fraction of molecular condensate ($\sim5\%$) is now sufficient to
explain the observed atom loss [8(3)\%].

\section*{Conclusions}
In conclusion, we have demonstrated that a
minimal mean-field model is sufficient to
qualitatively
explain a number of puzzling results in
Feshbach-resonant systems \cite{MAC02b}. Moreover, our refined
model \cite{MAC03} gives near-perfect quantitative agreement with the
double-pulse experiments \cite{DON02}, leaving little-to-no room for
additional loss mechanisms. Collapsing-condensate physics is therefore
understood as a matter of rogue dissociation, which leads to strong
losses in the threshold neighborhood, decreased remnant fraction for
decreasing interaction time---until very short times are reached,
and coherent remnant-burst oscillations. Ironically, the
Feshbach resonance has led to a regime dominated by
rogue dissociation, which apparently tends to counteract the production
of a molecular condensate.

\section*{Acknowledgements}
We acknowledgements Neil
Claussen and Eddy
Timmermans for helpful discussions, and the Academy of
Finland (MM and
KAS, projects 43336 and 50314), NSF (JJ, Grants
PHY-9801888 and
PHY-0097974), and NASA (JJ, Grant
NAG8-1428) for support.



\end{document}